\documentclass[10pt,reprint,aps,prc,showpacs]{revtex4-2}
\usepackage{graphicx,epstopdf,multirow,amsmath,rotating,enumitem,isotope,subfigure,csquotes}
\usepackage[breaklinks=true,colorlinks=true,linkcolor=blue,urlcolor=blue,citecolor=blue]{hyperref}

\begin{document}
%%%%%%%%%%%%% Specifying author and affiliation  %%%%%%%

 \author{Rinku Prajapat$^{1,2}$\footnote{Corresponding Author: R.KumarPrajapat@gsi.de}}

\affiliation{$^{1}$GSI Helmholtzzentrum für Schwerionenforschung GmbH, Darmstadt 64291, Germany}
\affiliation{$^{2}$Astronomy and Physics Department, Saint Mary's University, Halifax B3H C3C, Canada}

\title{Deciphering the influence of neutron transfer in Si-based fusion reactions around the Coulomb barrier}

	\date{\today}

%%%%%%%%%%%%%%%  ABSTRACT  %%%%%%%%%%%%%%%%%%%%%%%%%%%%%%%%%%%%%%%%%%%%%%%%%%%%%	

\begin{abstract}
	\hspace{-0.28cm}\textbf{Background:} The enhancement in sub-barrier fusion cross-sections caused by different intrinsic degrees of freedom, such as inelastic excitations and deformations, has been well explored recently. However, the influence of positive Q-value neutron transfer (PQNT) channels and their microscopic understanding on fusion dynamics have still been far from complete.  \\	
	\textbf{Purpose:} We aim to investigate the role of a few neutron transfer channels on the dynamics of fusion reactions around the Coulomb barrier by judicially selecting 11 different $^{28,30}$Si-induced systems. These reactions are chosen in such a way that they possess positive and negative Q-values for neutron transfer channels to make the comparison more apparent. Furthermore, a comparative study on fusion barrier parameters using different proximity potentials and parametrizations is also a prime goal.  \\
	\textbf{Method:} A channel coupling approach within the framework of a semiclassical model is being used to investigate the role of multi-neutron transfer with positive Q-values on fusion phenomena near and below the Coulomb barrier. The fusion barrier parameters have been extracted and analyzed within the framework of seven different potential models. \\
	\textbf{Results:} The sub-barrier fusion enhancement compared to the one-dimensional barrier penetration model (uncoupled) is investigated by considering collective excitations in colliding nuclei and multi-neutron transfer channels with Q $>$ 0 within the channel coupling model. Furthermore, GRAZING calculations are performed to predict the cross-section of target-like fragments after 2n pickup transfer. \\
   \textbf{Conclusion:} All the fusion excitation functions (EFs) have been successfully explained by the coupled channel calculations using the channel coupling model. Only the significant effect of up to 2n pickup transfer with Q $>$ 0 was found on sub-barrier fusion. Despite having positive Q values for transfer channels, no noticeable impact of more than 2n transfer was observed. GRAZING predictions are grossly in the same order as the quantitative contribution of 2n transfer channels observed by channel coupling model calculations. All potential models successfully explain the experimental barrier height and radius within 5 $\%$ and 20 $\%$ differences except for Prox 77 potential. Also, we have discussed the role of deformations after 2n transfer and its impact on sub-barrier fusion cross-section.

\end{abstract}


\maketitle

%%%%%%%%%%%%%%  Body of the paper %%%%%%%%%%%%%%%%%	

\section{\label{s1}Introduction}

Heavy-ion fusion and multinucleon transfer (MNT) reactions are two of the essential pathways to understand the diverse phenomena of nuclear physics, such as the production mechanism of exotic nuclei  and associated structural peculiarities, and stellar nucleosynthesis around the Coulomb barrier \cite{Commara2000, Watanabe2015,Tea2022}. In particular, fusion reactions are of special interest, governed by passing over or quantum penetration through the Coulomb barrier and influenced by various structural and dynamical processes. Regardless of the paramount importance of such reactions, their microscopic understanding has still been far from complete. For example, different degrees of freedom such as collective excitations (rotation or vibration), static deformations, and neutron transfer couplings distribute the single barrier into multiple fusion barriers, which give rise to enhancement in sub-barrier fusion cross-sections \cite{Jiang2021,Back2014,Beckerman1988,Dasgupta1998,Chauhan2020,Prajapat2022,Deepak2021,Stefanini2021,Tripathi2001}. Such dramatic process of enhancement is somewhat explored within the framework of collective excitations and deformations \cite{Jiang2021,Back2014,Beckerman1988,Dasgupta1998}. However, the role of neutron transfer with positive Q-value is not entirely understood due to the complexity of considering full-fledged transfer channel coupling in coupled channel calculations.

Thus, a series of theoretical calculations \cite{Esbensen2007,Zagrebaev2007,Zagrebaev2004,Umar2012} and experimental measurements \cite{Scarlassara2000,Stefanini2013,Jiang2014,Vandenbosch1997,Kolata2012,Henning1987, Prajapat2023,Kaur2024} have been performed to understand the impact of PQNT channels on sub-barrier fusion cross-sections. For instance, for the first time, Beckerman \textit{et al.} \cite{Beckerman1980} demonstrated the fusion enhancement caused by PQNT in $^{58,64}$Ni+$^{58,64}$Ni reactions around the Coulomb barrier energies. Later, a set of experiments proclaimed the role of PQNT on sub-barrier fusion enhancement. However, no such enhancement due to the PQNT channel was witnessed in multiple reactions, including $^{30}$Si+$^{156}$Gd \cite{Prajapat2022}, $^{32}$S+$^{112,116}$Sn \cite{Tripathi2001}, $^{132}$Sn+$^{58}$Ni \cite{Kohley2011}, $^{18}$O+$^{74}$Ge \cite{Jia2012}, despite having the PQNT channel, indicating the necessity of seminal experimental and theoretical understanding in this domain.

At the onset of theoretical understanding, Stelson \textit{et al.} \cite{Stelson1988,Stelson1990} followed by Rowely \textit{et al.} \cite{Rowley1992}, proposed a pragmatic approach by introducing the phenomenological models to incorporate the flow of neutrons between the colliding nuclei and their co-relation with sub-barrier fusion enhancement. Neutron flow occurs in a short interaction time of 10$^{-22}$ s. In his study, Rowley \textit{et al.}\cite{Rowley1992} found that neutron transfer with a negative Q-value broadens the barrier distribution (BD), which builds up neck formation between fusing participants. However, the same study also evidenced the antinecking configuration with the PQNT channel. Later, the quantum mechanical coupled channel (QCC) approach emerged to explain the dramatic nature of fusion enhancement. Nonetheless, it has been realized that incorporation of neutron transfer in QCC is a complex job owing to their consideration in the total coupled channel Schrödinger equations for a decomposition of the total wave function \cite{Zagrebaev2007,Zagrebaev2004}. Therefore, a strong motivation emerged to develop such an approach that can consider inelastic excitations as well as multi-neutron transfer in its calculation kernel. Because of this, an empirical channel coupling (ECC) approach \cite{Zagrebaev2001} was proposed, allowing both collective excitations and multi-neutron transfer with Q $>$ 0 simultaneously. This approach is well-pronounced and has recently been used in several near-barrier fusion studies with neutron rearrangement \cite{Prajapat2023,Prajapat2022,Zhang2010,Khushboo2019,Zagrebaev2001,Adel2012}. Apart from such studies, other reaction phenomena such as complete-incomplete fusion \cite{Prajapat2020_LiY,Prajapat2020_LiZr,Prajapat2021_6Li+Y,DKumar2017} and pre-equilibrium emission \cite{Prajapat2020_PEQ} are also emerging in recent time using light-heavy-ion induced reactions around the Coulomb barrier.

In addition, sincere efforts have been made to populate neutron-rich heavy nuclei via the MNT approach since fission and fragmentation reactions have very low production cross sections \cite{Zagrebaev2008,Son2023,Watanable2015}. This is also a fundamental crux of upcoming/existing Radioactive Ion Beams (RIBs) facilities \cite{Kubo2016}, more likely FRS/SFRS at GSI/FAIR in Germany, RIBF at RIKEN, and NSCL at MSU.
In the past, semi-classical GRAZING and dinuclear system (DNS) models have been successfully utilized to describe the cross-section, mass, and charge distribution of projectile/target-like fragments (PLFs/TLFs) from MNT reactions \cite{Mijatovic2016,Galtarossa2018,Adamian2010}. Despite these models' success, several measurements differ significantly from their predictions \cite{Welsh2017}. Hence, having the theoretical input using GRAZING before conducting such large-scale experiments is always desirable.

Different theoretical model approaches in heavy-ion collisions play a prominent role in extracting the fusion observables, such as barrier height and radius. Such parameters are also essential for synthesizing the superheavy elements and availability of RIBs worldwide \cite{Ichikawa2005}. To understand such intriguing interactions, different parametrized models and proximity potentials such as prox 2000, prox 2010 \cite{Gharaei2019,Dutt2010,Ghodsi2013}, and Bass 1980 \cite{Bass1977}, Kumari \textit{et al.} \cite{Kumari2015}, Zhang \textit{et al.} \cite{Zhang2016} become available in recent times and need attention to benchmark the data.

The facts mentioned above suggest that robust theoretical calculations are needed to investigate the role of a few neutron transfer channel couplings with Q $>$ 0 in addition to collective excitations of colliding nuclei. For this purpose, eleven different reactions are selected with the following motivations: (i) each pair of reactions has positive ($^{28}$Si+$^{62,64}$Ni, $^{30}$Si+$^{58}$Ni, and $^{28}$Si+$^{92,94,96}$Zr) and negative ($^{28}$Si+$^{58}$Ni, $^{30}$Si+$^{62,64}$Ni, and $^{28}$Si+$^{90}$Zr) Q-value transfer channels so that the impact of PQNT channels can be disentangled using the ECC model, (ii) projectiles are kept similar ($^{28,30}$Si) in order that structural effect of targets and PQNT couplings ($^{28}$Si $\longleftrightarrow$ $^{30}$Si after 2n transfer) can be studied, (iii) GRAZING calculations will be performed for those reactions which have the evidence of neutron transfer, and (iv)fusion barrier parameters will be extracted from available data and their comparison with a different theoretical model to deepen the current understanding.

The paper is organized as follows: Sec. \ref{s2} presents the theoretical formalism; results and interpretations are discussed in Secs.  \ref{s3} to \ref{s5}. Finally,  Sec. \ref{s6} concludes the work.

\begin{figure}
	%		\centering
	\includegraphics[width=8.2 cm]{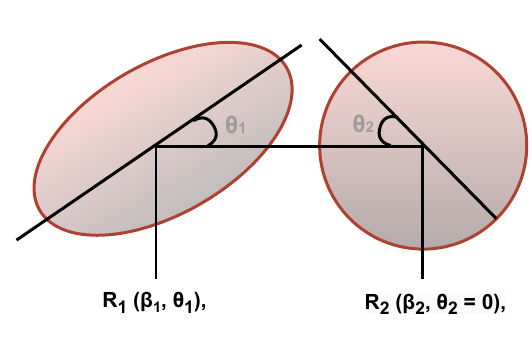}
	\caption{\label{FIG-1} Schematic of the two colliding nuclei: one is deformed, and the other is spherical in the reaction plane (see text for more details).}
\end{figure}

\section{\label{s2}Theoretical formalism} 

The fusion of two atomic nuclei at sub-barrier energies is governed by quantum tunneling phenomena \cite{Balantekin1998} and regulated by the real interaction potential, which consists of Coulomb, nuclear, and centrifugal terms. For low energy reactions in the light/intermediate-mass region, fusion cross-section ($\sigma_{fus}$), at a particular center of mass energy ($E_{c.m.}$) and angular momentum ($\ell$) can be expressed as the sum of all partial waves [Eq. \ref{eq1}]:

\begin{figure}
	%		\centering
	\includegraphics[width=8.8 cm]{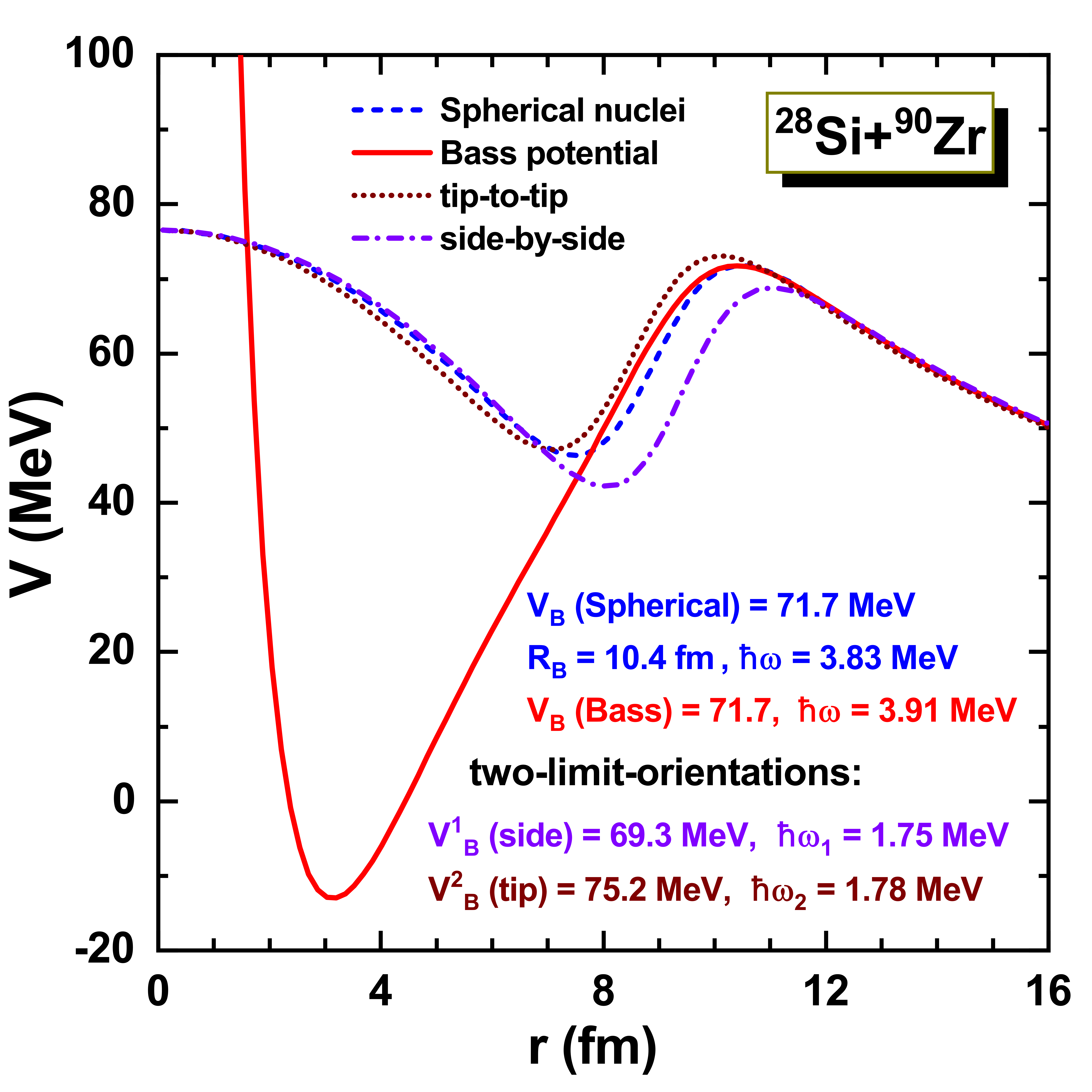}
	\caption{\label{FIG-2} Total interaction potential (V) as a function of the interaction radius (r) for $^{28}$Si+$^{90}$Zr reaction (see text for details).}
\end{figure}

\begin{equation}
	\label{eq1}
	\sigma_{fus}(E_{c.m.}) = \pi \lambdabar^{2}\sum_{\ell=0}^{{\infty}}(2\ell+1) T_{\ell}(E_{c.m.})
\end{equation}

The absorption probability of $\ell^{th}$-partial wave $T_{\ell}^{HW}$($E_{c.m.}$)=[1+exp\{(2$\pi$/$\hbar$$\omega_{\ell}$)($V_{b\ell}$--$E_{c.m.}$)\}]$^{-1}$  has been determined using Hill-Wheeler approach \cite{HillWheeler1953}. Here, $V_{b\ell}$ and  $\hbar\omega_{\ell}$ are the barrier height (in MeV) and curvature for  $l^{th}$ partial wave, respectively. However, the curvature and radius ($R_{bl}$) are independent of angular momentum in case of $\ell$=0. In such scenario, they can be followed as the $S$-wave values $\hbar\omega_{l}$=$\hbar\omega$, and $R_{bl}$=$R_{b}$.  Therefore, the fusion barrier can be extracted by fitting the measured fusion cross section with Wong's formula \cite{Wong1973} (Eq. \ref{eq2})

\begin{equation}
	\label{eq2}
	\sigma_{fus}(E_{c.m.}) = \frac{R_{b}^{2}\hbar \omega}{2E_{c.m.}}ln\Big\{1+exp\Big[\frac{2\pi}{\hbar \omega} \Big(E_{c.m.}- V_{b}\Big)\Big]\Big\}
\end{equation}

At the energies well above the Coulomb barrier, i.e. ($E_{c.m.}-V_{b}$)$\geq$ $\hbar \omega$/2$\pi$, Eq. \ref{eq2} can be approximated to a simplified classical expression as follows [Eq. \ref{eq3}]

\begin{equation}
	\label{eq3}
	\sigma_{fus}(E_{c.m.}) = \pi R_{b}^{2}\Big(1-\frac{V_{b}}{E_{c.m.}}\Big),
\end{equation}
The linear dependence of $\sigma_{fus}$ on $1/E_{c.m.}$ is already known at the above barrier energies.
From the linear fit of the data, the values of $V_{b}$ and $R_{b}$ can be calculated for the different systems to compare with those obtained from the different potential models (see Sec. \ref{s5}).

But, the interaction potential not only depends on the relative separation between the colliding partners but also on their deformation characteristics ($\beta_{p}$, $\beta_{t}$) and mutual orientations ($\theta_{p}$, $\theta_{t}$), here p and t refers to projectile and target nucleus, respectively. Hence, the potential energy can be written as follows \cite{Zagrebaev2004}:

\begin{equation}
	\begin{split}
		\label{eq4}
		%	\centering
		V_{p,t}(r, \beta_{p}, \beta_{t}, \theta_{p}, \theta_{t}) =  V_{C}(r, \beta_{p}, \beta_{t}, \theta_{p}, \theta_{t})  \\ +  V_{N}(r, \beta_{p}, \beta_{t}, \theta_{p}, \theta_{t})  + \frac{1}{2}C_{p}(\beta_{p}-\beta_{p}^{0})^{2} + \frac{1}{2}C_{t}(\beta_{t}-\beta_{t}^{0})^{2}
	\end{split}
\end{equation}

\begin{figure}
	%		\centering
	\includegraphics[width=9.5 cm]{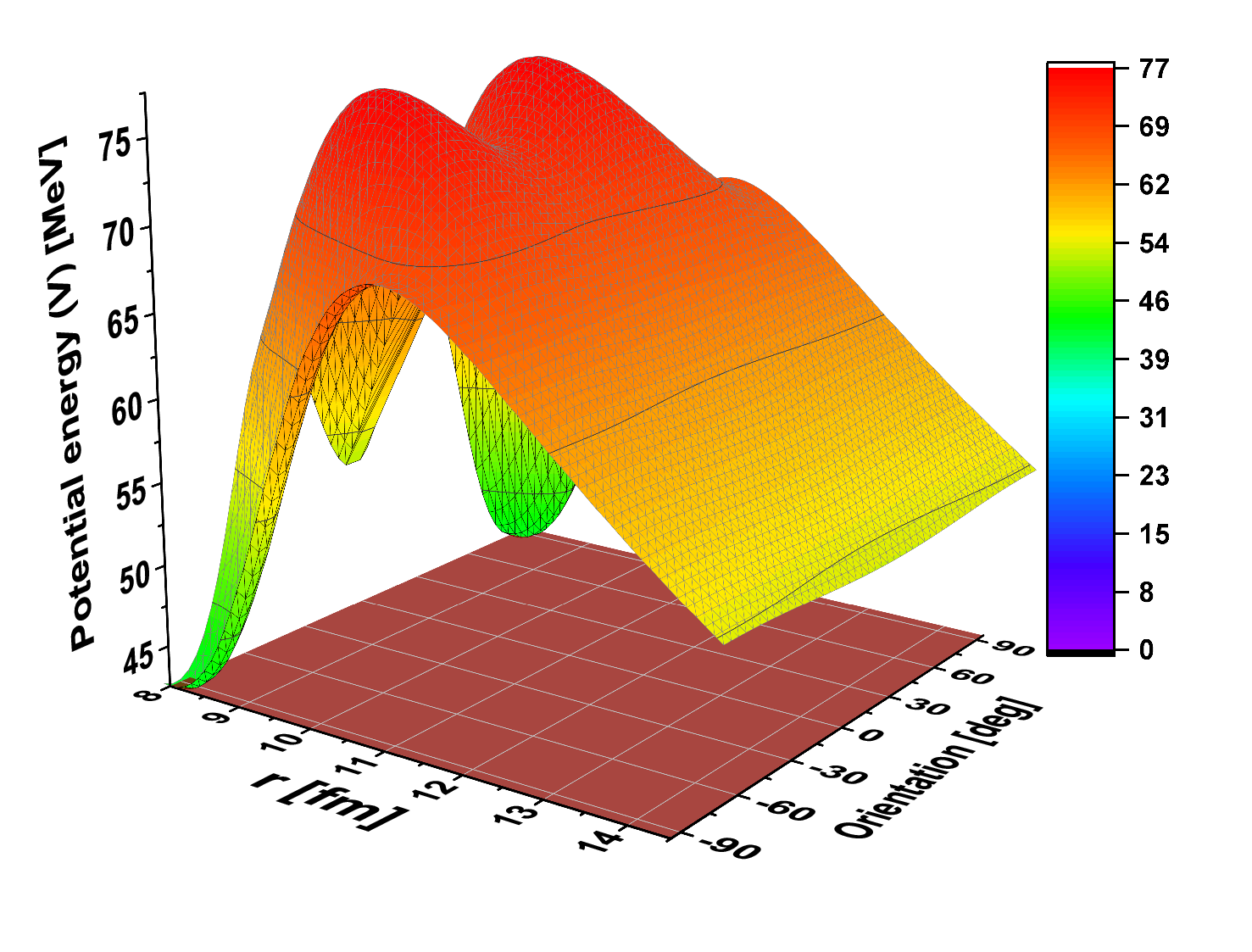}
	\caption{\label{FIG-3} Total interaction potential (V) as a function of distance (r) and relative orientation for $^{28}$Si ($\beta_{2}$ = -0.407, $\beta_{4}$ = +0.25) and $^{90}$Zr nuclei.}
\end{figure}

\begin{figure*}
	%		\centering
	\includegraphics[width=18.0 cm]{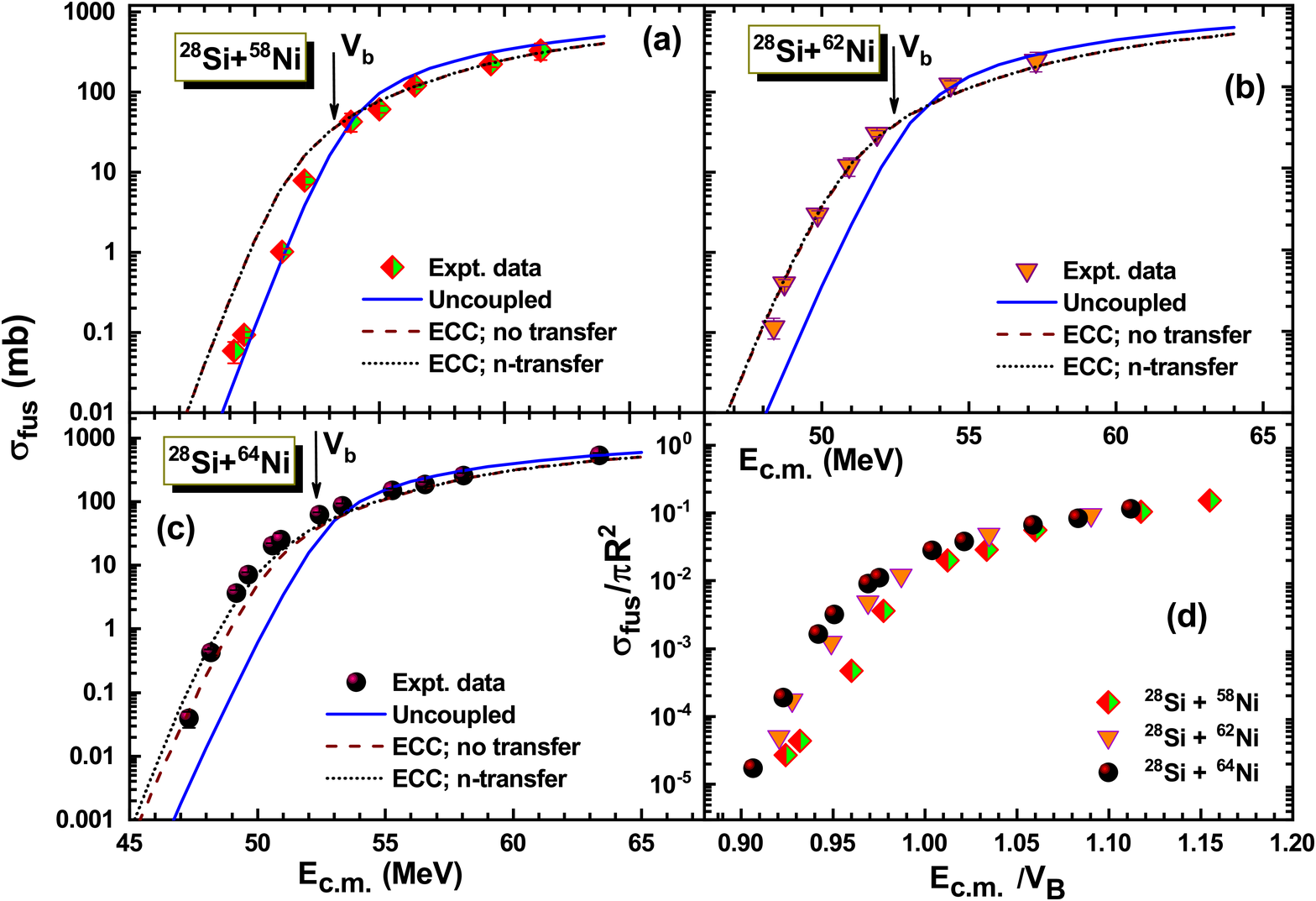}
	\caption{\label{FIG-4}Comparison between the experimental fusion data \cite{Stefanini1984} and theoretical calculations using the empirical channel coupling model by considering no coupling (uncoupled),  without neutron transfer, and neutron transfer into account for fusion reactions of (a) $^{28}$Si+$^{58}$Ni, (b) $^{28}$Si+$^{62}$Ni, and (c) $^{28}$Si+$^{64}$Ni.  Figure (d) represents the comparison between $^{28}$Si+$^{58,62,64}$Ni on reduced scales.}
\end{figure*}

\begin{figure}
	%		\centering
	\includegraphics[width=8.8 cm]{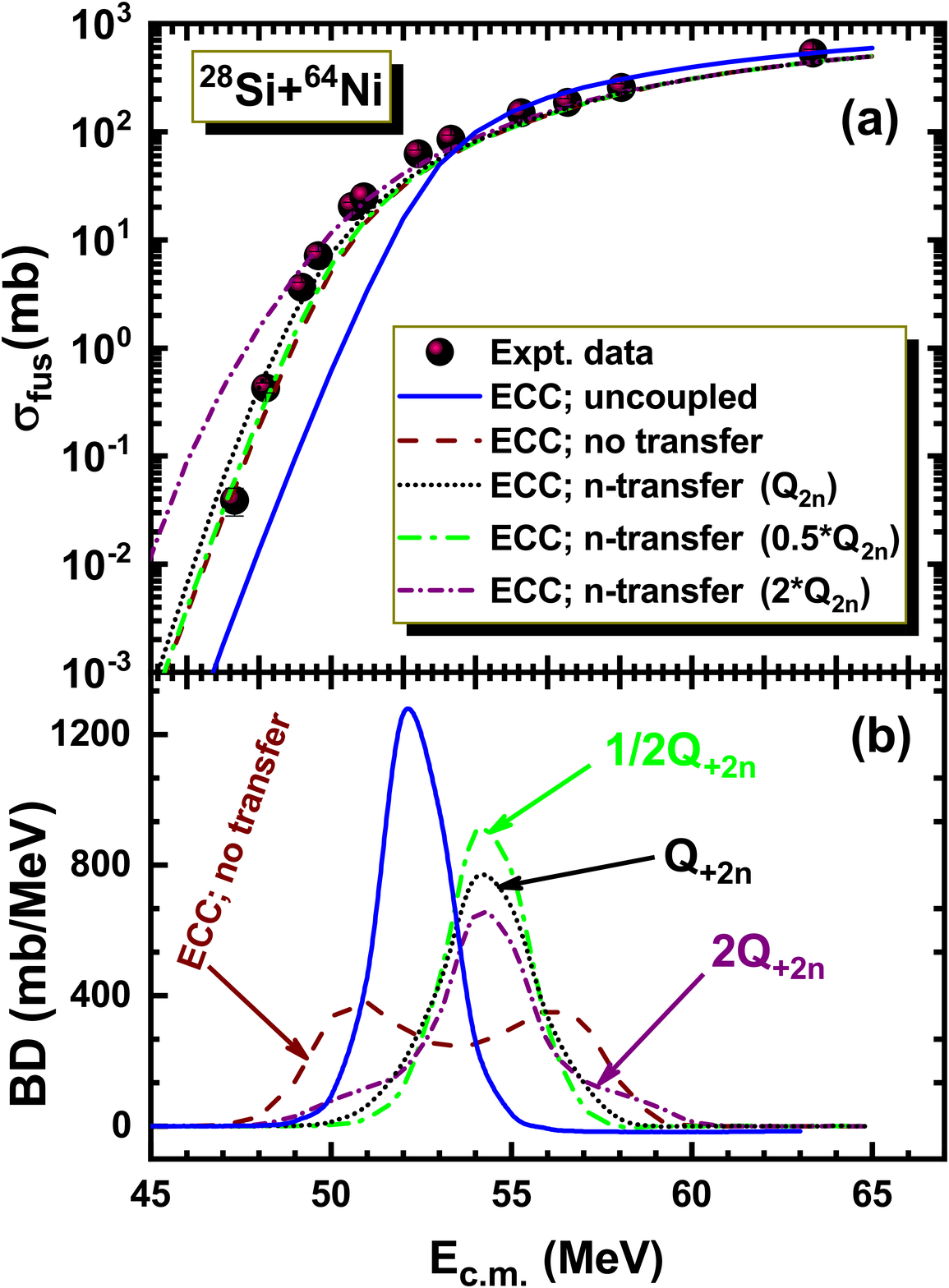}
	\caption{\label{FIG-5} (a) Fusion excitation function for  $^{28}$Si+$^{64}$Ni \cite{Stefanini1984} reaction and comparison with ECC model calculations by accounting neutron transfer with real $Q_{+2n}$ value and by assuming the $Q_{+2n}$ values as halved and doubled (see text), (b) corresponding barrier distribution for $^{28}$Si+$^{64}$Ni reaction using ECC model predictions. }
\end{figure}

The parameters $\beta_{p, t}$, $\beta_{p, t}^{0}$, and $C_{p,t}$ are the dynamic and static quadrupole deformation, and stiffness parameters of projectile-target nuclei, respectively. The stiffness is calculated within the framework of the liquid drop model. Here, $\theta_{p, t}$ are the orientations of the symmetry axes, as shown in Fig. \ref{FIG-1}.  
In such a scenario, mainly there are two cases of potential energy such as (i) the interaction potential for two spherical nuclei ($\beta_{p, t}$ = 0) is almost close to the Bass barrier, (ii) for deformed nuclei, the potential barrier will be calculated at different relative orientations. However, for representation purposes, only two limiting cases of $\theta_{p, t}$ =$\pi$/2 (side-by-side orientation) or $\theta_{p, t}$ = 0 (tip-to-tip orientation) are shown in Fig. \ref{FIG-2}. Furthermore, a multidimentional character of potential barrier for $^{28}$Si+$^{90}$Zr($^{28}$Si; deformed and $^{90}$Zr; spherical) is shown in Fig. \ref{FIG-2}. Similarly, the two-dimensional interaction potential for the deformed nucleus $^{28}$Si ($\beta_{2}$ = -0.407, $\beta_{4}$ = +0.25) and the spherical nucleus $^{90}$Zr is shown in Fig. \ref{FIG-3} as a function of distance (r) and relative orientation.
One can observe the difference in different barrier observables (height, position, and curvature) from the same figure (Fig. \ref{FIG-2}) influencing the fusion cross-sections. 
To simulate such multidimensional character of potential barriers, one has to solve a multidimensional Schr$\ddot{o}$dinger equation.

Within the ECC model, for the spherical nuclei which depend only on the single degree of freedom, coupling to their surface vibrations, the transmission probability [$T_{\ell}(E_{c.m.}$)] is calculated by averaging the barrier height $V_{b}$ (Eq. \ref{eq5})

\begin{equation}
	\label{eq5}
	T_{\ell}(E_{c.m.}) = \int f(V_{b})  T_{\ell}^{HW}(V_{b}; E_{c.m.}) dV_{b}
\end{equation}

where $f(V_{b})$ is the barrier distribution function \cite{Zagrebaev2001} and can be estimated using the normalization condition $\int f(V_{b})dV_{b}$ = 1.

Meanwhile, the sub-barrier fusion cross-section and associated dynamics of spherical and statistically deformed shape nuclei mainly depend on the coupling of their surface vibrations and the mutual orientation of colliding partners. Hence, the penetration probability is average over the deformation-dependent barrier height for the deformed nuclei, as follows [Eq. \ref{eq6}]:

\begin{equation}
	\begin{split}
		\label{eq6}
		T_{\ell}(E_{c.m.}) = \frac{1}{4}\int_{0}^{\pi} \int_{0}^{\pi}   T_{\ell}^{HW}(V_{b}(\beta_{p,t}; \theta_{p,t}), E_{c.m.}) \\ \times sin\theta_{1} sin\theta_{2} d\theta_{1} d\theta_{2}
	\end{split}
\end{equation}

To contemplate the multi-neutron transfer (rearrangement) with Q $>$ 0, where incoming flux may penetrate the multidimensional
Coulomb barrier for the different neutron transfer channels, the penetration probability was calculated by Eq. \ref{eq5} or \ref{eq6} in which $T_{\ell}^{HW}$ has to replace by the following expression [Eq. \ref{eq7}]:

\begin{equation}
	\begin{split}
		\label{eq7}
		T_{\ell}^{HW}(V_{b}; E_{c.m.}) =  \frac{1}{N_{tr}}  \sum_{x = 0}^{4}  \int_{- E_{c.m.}}^{Q_{xn}} \alpha_{k}(E_{c.m.}, \ell, Q) \\ \times T_{\ell}^{HW}(V_{b}; E_{c.m.} + Q)dQ
	\end{split}
\end{equation}

$N_{tr}$ and $Q_{xn}$ are the normalization constant and Q value of $x$ neutron transfer from the ground state of one participant to the ground state of other participating nuclei. Then, the probability of $x$ neutron transfer with Q $>$ 0 can be estimated using the following expression [Eq. \ref{eq8}]:

\begin{equation}
	\label{eq8}
	\alpha_{k}(E_{c.m.}, \ell, Q) = N_{k} e^{-C(Q-Q_{opt})^{2}} e^{-2\delta(D(E_{c.m.}, \ell) - D_{0})}
\end{equation}

where $D(E_{c.m.}, \ell)$ and $D_{0}$ are the distance of closest approach of two colliding partners and $d_{0}(A_{p}^{1/3} + A_{t}^{1/3})$, respectively with $d_{0}$ = 1.2 fm \cite{Oertzcn1987} and $A_{p}, A_{t}$ are the mass number of projectile and target nuclei. The parameter $\delta$ is $\delta$ = $\delta(\epsilon_{1})$ + $\delta(\epsilon_{2})$ + $\delta(\epsilon_{3})$ +...+ $\delta(\epsilon_{k})$, for the sequential transfer of k neutrons with $\delta(\epsilon_{k}$) =  $\sqrt{2\mu_{n}\epsilon_{k}/\hbar^{2}}$, where $\epsilon_{k}$ is the sepration energy of the k-th transfered neutron. The parameters C = $R_{b}\mu_{p,t}$/4$\delta \hbar^{2} V_{b}$ with $\mu_{p,t}$ as the reduced mass of projectile and target and $Q_{opt}$ is the optimum Q value for nucleon transfer. The nucleon transfer in heavy-ion-induced reactions is regulated mainly by the $Q_{opt}$. However, the charge of colliding partners doesn't change after neutron transfer channels, leading to $Q_{opt}$ = 0 \cite{Henning1978}.

 Thus, the probability for k neutron transfer at  $E_{c.m.}$ and  $\ell$, in the entrance channel to the final state with Q $\leq$ Q$_{0}(k)$, where Q$_{0}(k)$ is the Q-value for the ground state to ground state transfer reaction, can be as written as follow [Eq. \ref{eq9}]: 

\begin{equation}
	\label{eq9}
	\alpha_{k}(E_{c.m.}, \ell, Q) = N_{k} e^{{-CQ}^{2}} e^{-2\delta(D(E_{c.m.}, \ell) - D_{0})}
\end{equation}

where $N_{k}$ is expressed as :

\begin{equation}
	\label{eq10}
	N_{k}(E_{c.m.}) = \Big[ \int_{-E_{c.m.}}^{Q_{0}(k)}  exp(-CQ^{2})dQ\Big]^{-1}
\end{equation}

However, $N_{k}$ and second exponent of Eq. \ref{eq8} should be replaced by 1 if $D(E_{c.m.}$, $\ell$) $<$ $D_{0}$.

As one can notice from Eq. \ref{eq7}, the gain in energy due to positive Q-value neutron transfer channels may lead to enhancement in fusion probability at the sub-barrier energy domain. However, it is to be noted that up to four neutron transfer channels are taken into account in present calculations as no significant effect can be visible for 5n and 6n transfer.

\begin{table}
	\caption{ The Woods-Saxon ion-ion nuclear potential parameters with Aky\"{u}z-Winther (AW) parametrization; well depth ($V_{0}$), radius parameter ($r_{0}$), diffuseness ($a$), and Bass barier curvature ($\hbar \omega_{b})$. }
	\label{table1}
	\begin{ruledtabular}
		\begin{center}
			\renewcommand{\arraystretch}{1.0}
			\begin{tabular}{ccccccc}
				Reaction & $V_{0}$ (MeV) & $r_{0}$ (fm) & a (fm) & $\hbar \omega_{b}$ (MeV)\\
				\hline
				
				$^{28}$Si + $^{58}$Ni & -62.1 & 1.174 & 0.652 & 3.68  \\
				$^{28}$Si + $^{62}$Ni & -62.9 & 1.174  & 0.653 & 3.61  \\
				
				$^{28}$Si + $^{64}$Ni & -63.3 & 1.174 & 0.654 & 3.56  \\
				
				$^{28}$Si + $^{90}$Zr  & -67.1 & 1.176 & 0.661 & 3.83 \\
				
				$^{28}$Si + $^{92}$Zr & -67.3 &  1.176 & 0.662 & 3.78  \\
				
				$^{28}$Si + $^{94}$Zr & -67.6 & 1.176 & 0.662 & 3.76  \\
				
				$^{28}$Si + $^{96}$Zr & -67.8 &  1.176 & 0.662 & 3.76 \\
				
				$^{28}$Si + $^{144}$Nd & -72.3 & 1.178 & 0.67 & 4.05 \\
				
				$^{30}$Si + $^{58}$Ni & -62.9 & 1.174 & 0.654 & 3.56   \\
				
				$^{30}$Si + $^{62}$Ni & -63.2 & 1.175 & 0.655 & 3.49   \\
				
				$^{30}$Si + $^{64}$Ni & -63.3 & 1.175 & 0.656 & 3.44  \\
				
			\end{tabular}
		\end{center}
	\end{ruledtabular}
\end{table}

\section{\label{s3}Results and discussion} 

The calculations have been performed for different judicially selected systems using the empirical channel coupling approach \cite{Zagrebaev2001}, which considers the couplings to collective states of colliding nuclei and multi-neutron transfer channels with Q $>$ 0. To perform such calculations, the standard Woods-Saxon ion-ion nuclear potential with Aky\"{u}z-Winther (AW) parametrization has been used. The AW parameters,  well depth ($V_{0}$), radius parameter ($r_{0}$), diffuseness ($a$), and Bass barier curvature ($\hbar \omega_{b})$ are listed in Table \ref{table1}. The parameters of the vibrational excitations are taken from the NRV experimental database \cite{NRV}. Furthermore, the other parameter necessary for ECC calculation is stiffness, which was taken into account from the liquid drop model \cite{Bohr1998}. The deformation parameters for the rotational excitation in $^{28}$Si are considered from Ref. \cite{Kalkal2010}. A detailed discussion of each selected system is as follows.

\begin{figure*}
	%		\centering
	\includegraphics[width=18.0 cm]{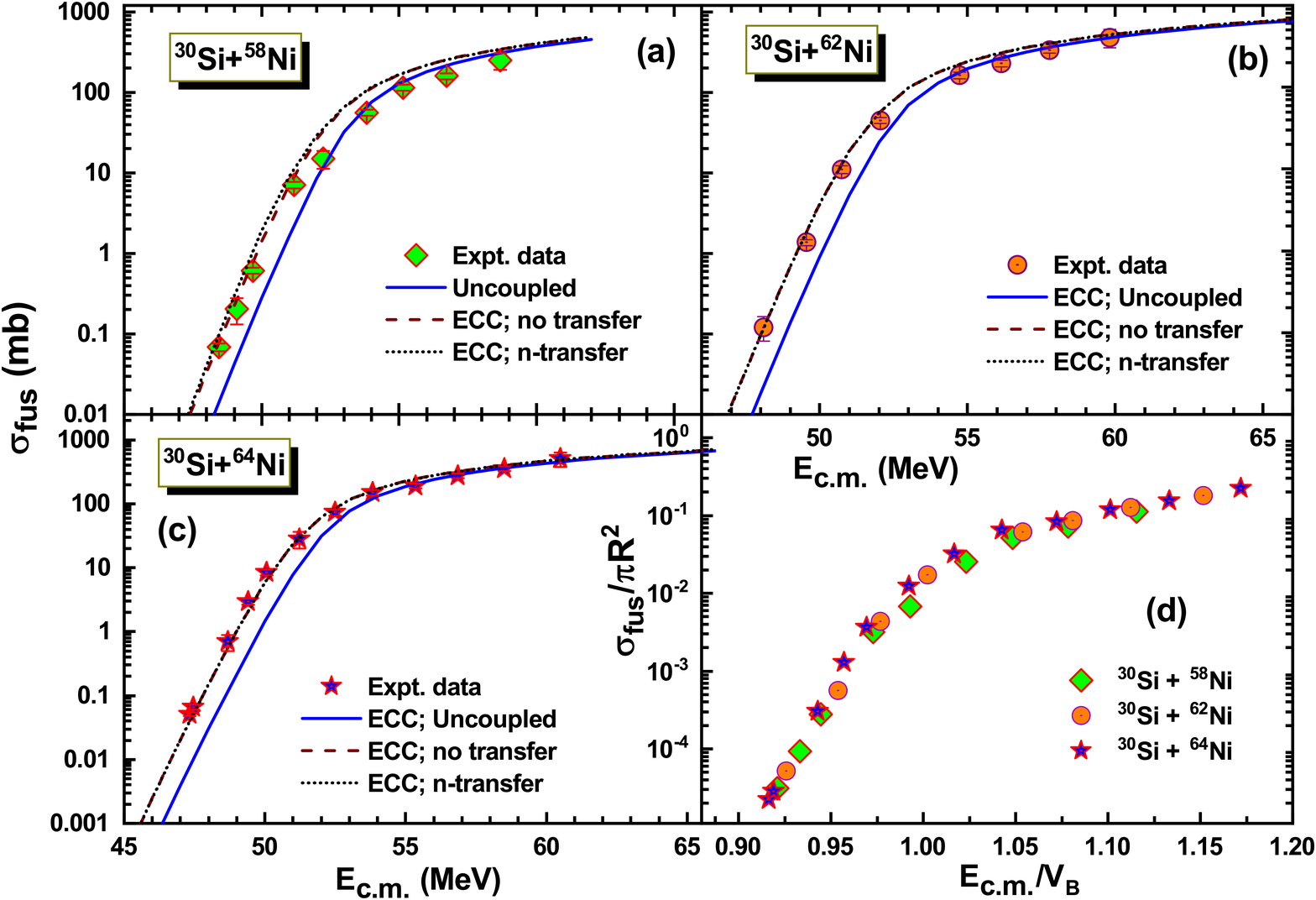}
	\caption{\label{FIG-6} Same as Fig. \ref{FIG-4} for $^{30}$Si-induced reactions on Ni isotopes. The experimental data is taken from Ref. \cite{Stefanini1984}.}
\end{figure*}

\subsection{$^{28}$Si+$^{58,62,64}$Ni}

Figure \ref{FIG-4}(a, b, c) depicts the comparison between the theoretical predictions with uncoupled, channel coupling (with and without neutron transfer), and the experimental fusion excitation functions for $^{28}$Si+$^{58,62,64}$Ni \cite{Stefanini1984}, respectively. Stefanini \textit{et al.} \cite{Stefanini1984} hinted that the cause of enhancement may be due to the presence of the PQNT channel and sought to perform rigorous theoretical calculations to confirm the same. Therefore, we have performed robust theoretical calculations to deepen the underlying mechanism. For theoretical calculations, $^{28}$Si is considered as pure rotor with quadrupole deformation $\beta_{2}$ = -0.407 and hexapole deformation $\beta_{4}$ = +0.25 \cite{Kalkal2010} whereas vibrational couplings with stiffness parameters C = 7.149 ($^{58}$Ni), 7.164 ($^{62}$Ni), and 7.127 ($^{64}$Ni) are considered in all the target isotopes of Ni. One can notice that ECC uncoupled (without any coupling) calculations are in reasonable agreement with experimental data below the Coulomb barrier V$_{b}$ (arrow in Fig. \ref{FIG-4}(a)) whereas slightly overpredicting the experimental data at above barrier energies. Further inclusion of empirical couplings in projectile-target nuclei along with and without neutron transfer channel overestimates the experimental data at below barrier energies, whereas explaining it quite well at above barrier energies. The predicted cross-sections are slightly higher than the experimental data at belowe barrier region, possibly due to the rotational couplings in projectile and mutual excitations in projectile-target nuclei.
However, the ECC predictions are almost identical when we consider with and without neutron transfer couplings for $^{28}$Si+$^{58}$Ni, which might be due to the absence of any PQNT channels (see Table \ref{table2}). It is imperative to mention that $^{28}$Si+$^{58}$Ni possesses negative Q values for all the neutron transfer channels (up to 6n), as can be observed from Table \ref{table2}. Therefore, even after consideration of neutron transfer channels along with channel couplings, no effect has been observed on fusion cross-sections, and the predicted cross-sections coincide with the channel couplings without neutron transfer curve. As a result, it can be concluded that transfer channels do not influence fusion cross-sections in $^{28}$Si+$^{58}$Ni system.

\begin{table*}
	\caption{Ground state $Q_{+xn (-xn)}$ (MeV) values for multi neutron pickup (by a projectile from target) and stripping transfer channels for $^{28, 30}$Si induced reactions. }
	\label{table2}
	\begin{ruledtabular}
		\begin{center}
			\renewcommand{\arraystretch}{1.0}
			\begin{tabular}{ccccccc}
				Reaction & $Q_{+1n(-1n)}$ & $Q_{+2n(-2n)}$  & $Q_{+3n(-3n)}$ & $Q_{+4n(-4n)}$ & $Q_{+5n(-5n)}$ & $Q_{+6n(-6n)}$ \\
				\hline			
				$^{28}$Si + $^{58}$Ni & -3.7 (-8.2) & -3.4 (-10.1) & -13.4 (-21.3) & -18.4 (-25.7) & -31.6 (-39.9) & -38.2 (-47.9)  \\
				$^{28}$Si + $^{62}$Ni & -2.1 (-10.3) & +0.7 (-14.0) & -4.1 (-26.9) & -3.9 (-33.0) & -11.6 (-48.2) & -14.4 (-58.1)  \\
				
				$^{28}$Si + $^{64}$Ni & -1.2 (-11.1) & +2.6 (-15.4) & -1.4 (-28.7) & -0.04 (-35.9) & -6.9 (-52.3) & -8.4 (-62.7)  \\
				
				$^{28}$Si + $^{90}$Zr  & -3.5 (-10.0) & -2.2 (-14.7) & -8.0 (-27.0) & -8.2 (-33.7) & -16.6 (-48.3) & -18.9 (-58.1)  \\
				
				$^{28}$Si + $^{92}$Zr & -0.2 (-10.4) & +3.3 (-15.5) & -2.1 (-28.1) & -2.2 (-35.2) & -10.1 (-50.7) & -12.0 (-62.0)  \\
				
				$^{28}$Si + $^{94}$Zr & +0.3 (-10.7) & +4.1 (-16.2) & +2.1 (-29.6) & +4.1 (-38.2) & -3.4 (-54.8) & -5.2 (-65.7)  \\
				
				$^{28}$Si + $^{96}$Zr & +0.6 (-11.6) & +4.8 (-18.5) & +3.1 (-33.1) & +5.6 (-41.3) & +1.5 (-57.5) & +1.8 (-68.7)  \\
				
				$^{28}$Si + $^{144}$Nd & +0.7 (-11.4) & +5.1 (-17.2) & +1.9 (-30.9) & +3.1 (-38.6) & -2.7 (-54.6) & -3.3 (-64.9)  \\
				
				$^{30}$Si + $^{58}$Ni & -5.6 (-1.6) & -6.7 (+1.3) & -18.8 (-8.1) & -25.5 (-10.7) & -40.7 (-23.0) & -48.8 (-28.3)  \\
				
				$^{30}$Si + $^{62}$Ni & -4.0 (-3.8) & -2.6 (-2.6) & -9.5 (-13.7) & -10.9 (-18.0) & -20.7 (-31.3) & -24.9 (-38.5)  \\
				
				$^{30}$Si + $^{64}$Ni & -3.1 (-4.5) & -0.7 (-4.0) & -6.8 (-15.4) & -7.1 (-20.9) & -16.0 (-35.4) & -18.9 (-43.1)  \\
				
			\end{tabular}
		\end{center}
\end{ruledtabular}\end{table*}

Similarly, the ECC calculations was performed for $^{28}$Si+$^{62,64}$Ni systems as shown in Fig. \ref{FIG-4}(b) and (c), respectively. One can observe that ECC predictions without considering neutron transfer are in excellent agreement with experimental fusion data for $^{28}$Si+$^{62}$Ni, whereas it was found to be
lower compared to the experimental data for $^{28}$Si+$^{64}$Ni. This means that the fusion data do not show any significant enhancement due to Q$_{+2n}$ = +0.7 MeV for $^{28}$Si+$^{62}$Ni system as compared to $^{28}$Si+$^{64}$Ni having Q$_{+2n}$ = +2.6 MeV. This could be due to the fact that Q$_{+2n}$ value is around four times larger for $^{28}$Si+$^{64}$Ni than $^{28}$Si+$^{62}$Ni reaction. This is also demonstrated theoretically using the ECC model calculations in Fig. \ref{FIG-5} for $^{28}$Si+$^{64}$Ni system. As if one halved the Q$_{+2n}$ value such as Q$_{+2n}$ = +2.6/2= +1.3 MeV, then the barrier distribution becomes slightly narrow and increases the barrier height (see Fig. \ref{FIG-5}(b)). As a result, enhancement in fusion cross-section is reduced and becomes much closer to ECC calculations without neutron transfer. Moreover, if the Q$_{+2n}$ value is doubled, such as Q$_{+2n}$ = 2$\times$(+2.6) = +5.2 MeV, broaden the BD and suppress the barrier height, as shown in the figure \ref{FIG-5}(b). This reflects the stronger effect on predicted sub-barrier fusion cross-sections and drastically increases the fusion enhancement, as shown in Fig. \ref{FIG-5}(a).

Furthermore, the $\sigma_{fus}$ and $E_{c.m.}$ was scaled by geometrical cross-section ($\pi$R$^{2}$) and Bass-barrier height (V$_{B}$), respectively, as shown in Fig. \ref{FIG-4}(d). By doing so, the effect of different barrier heights and positions can be removed to make the comparison more apparent and to avoid any other effect. The radius (R) was calculated as R = 1.2*(A$_{p}^{1/3}$+A$_{t}^{1/3}$), where A$_{p}$ and A$_{t}$ are the mass numbers of projectile and target, respectively. One can observe the considerable amount of enhancement for $^{28}$Si+$^{64}$Ni on the reduced scale as compared to $^{28}$Si+$^{58,62}$Ni reflecting the significant role of the PQNT channel in the first system.

\subsection{$^{30}$Si+$^{58,62,64}$Ni}

Figure \ref{FIG-6}(a, b, c) shows the comparison between experimental fusion EFs \cite{Stefanini1984} and those calculated using the empirical channel coupling model with no coupling (uncoupled), with and without neutron transfer channels for $^{30}$Si+$^{58,62,64}$Ni systems. For theoretical calculations, both projectile and target isotopes are considered as vibrators. It can be observed that ECC calculations without considering neutron transfer couplings successfully explain the experimental data throughout the energy domain for all three systems. However, no effect of fusion enhancement was observed when neutron transfer channels were taken into account, which might be due to the fact that all transfer channel Q -values are negative (see Table \ref{table2}) for $^{30}$Si induced reaction except the case for $^{30}$Si+$^{58}$Ni (Q$_{-2n}$ = +1.3). Furthermore, a comparison between all three systems on the reduced scale of $\sigma_{fus}$ and $E_{c.m.}$ shows the overlapping pattern in Fig. \ref{FIG-6}(d), revealing no effect of 2n stripping transfer channel in $^{30}$Si+$^{58}$Ni system despite having PQNT channel.

One can conclude from the facts mentioned above that the availability of transfer channels with positive Q-values for $^{28}$Si+$^{62, 64}$Ni is readily possible and explained based on ECC model calculations by considering two neutron pickup transfer ($^{28}$Si $\rightarrow$ $^{30}$Si). However, a similar approach cannot apply for $^{28}$Si+$^{58}$Ni despite having a positive Q-value for two neutron stripping ($^{30}$Si $\rightarrow$ $^{28}$Si). Similary no significient effect of PQNT has been reported in $^{30}$Si+$^{156}$Gd \cite{Prajapat2022} system despite of having low Q$_{+2n}$ = +0.8 MeV using ECC model.

\begin{figure*}
	%	\begin{subfigure}
		%		\centering
		\includegraphics[width=18.0cm]{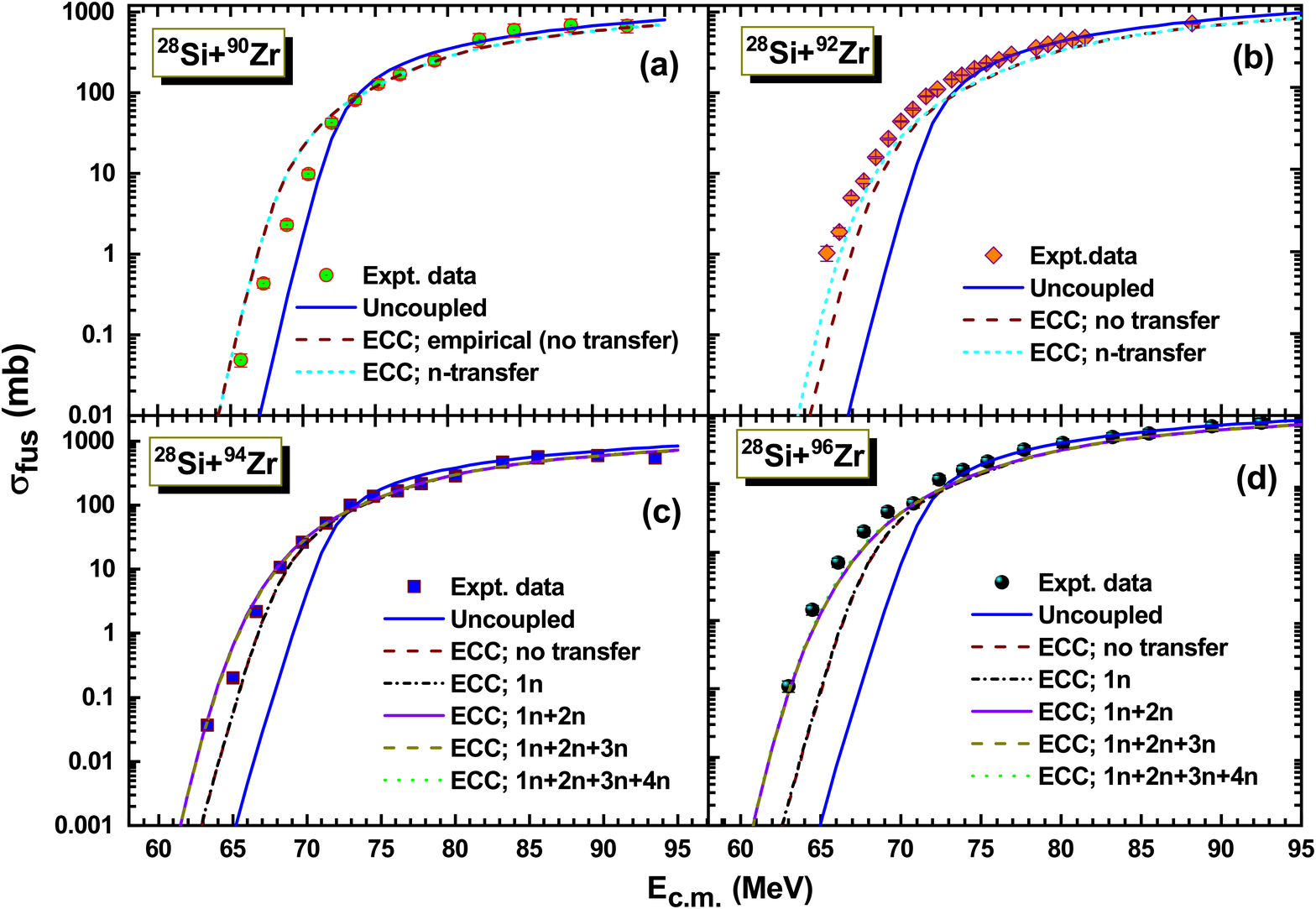}
		%	\end{subfigure}
	\caption{\label{FIG-7}  Same as Fig. \ref{FIG-5} for $^{28}$Si+$^{90, 92, 94, 94}$Zr systems (see text for the details). The experimental data is taken from Refs. \cite{Kalkal2010,Khushboo2017}.}	
\end{figure*}

\subsection{$^{28}$Si+$^{90, 92, 94, 94}$Zr}

In order to disentangle the role of the PQNT channels on sub-barrier fusion dynamics, $^{28}$Si+$^{90, 92, 94, 94}$Zr systems \cite{Kalkal2010,Khushboo2017} are selected, which are the admixture of positive and negative ground state Q-value for neutron transfer channels, as can be seen in Table \ref{table2}. It is good to mention that it has been pointed out in Refs. \cite{Kalkal2010,Khushboo2017} that robust theoretical calculations incorporating the multi-neutron transfer channels are needed, as the coupled channel calculations failed to explain the sub-barrier fusion data. Therefore, we present detailed calculations using the ECC model, which can consider both the multi-neutron transfer channels and inelastic excitations in colliding nuclei. Rotational couplings in $^{28}$Si and vibrational couplings in target nuclei are being adopted for theoretical calculations.
Figure \ref{FIG-7}(a) shows the comparison between experimental data \cite{Kalkal2010} and predicted values using the no coupling (uncoupled), ECC without neutron transfer, and ECC with neutron transfer for the $^{28}$Si+$^{90}$Zr system (transfer Q values are negative). The apparent signature of Fig. \ref{FIG-7}(a) is the significant enhancement in the sub-barrier fusion cross-sections as compared to no coupling limit (uncoupled). Another striking feature is that ECC calculations without neutron transfer slightly overpredict the experimental fusion cross-sections below the Coulomb barrier, whereas explaining the data nicely at above barrier energies. 
However, the scenario is different for the case of $^{28}$Si+$^{92}$Zr system having Q$_{+2n}$ = +3.3 MeV. Thus, ECC with neutron transfer predictions has a larger enhancement than ECC with no transfer due to the presence of Q$_{+2n} >$ 0 (see Fig. \ref{FIG-7}(b)). Moreover, these predictions slightly underpredict the experimental data and need a deeper understanding.
One can say that the fusion enhancement due to neutron transfer is more minor for systems with a weak coupling impact to collective states.

\begin{figure}
	\includegraphics[width=8.8cm]{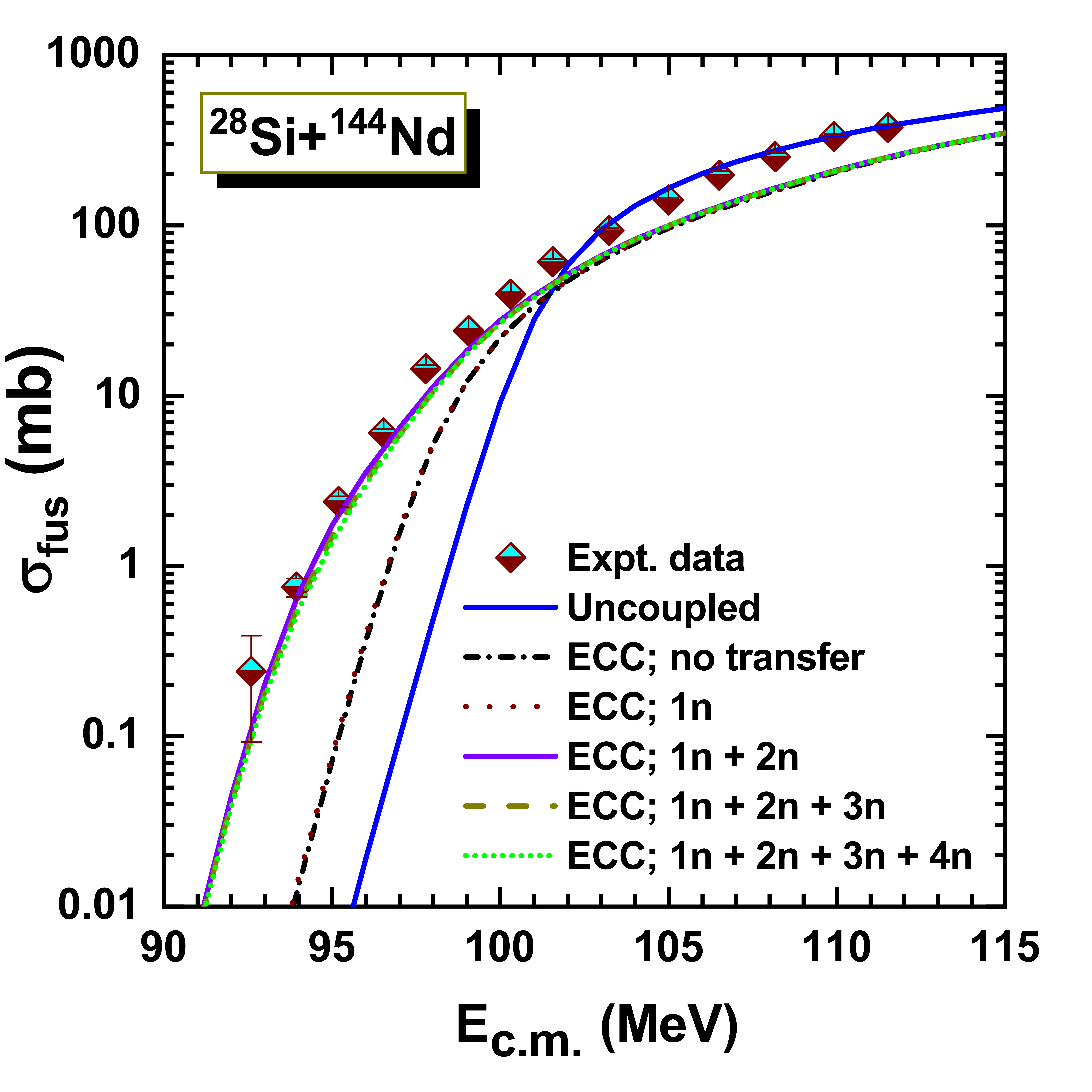}
	\caption{\label{FIG-8} Same as Fig. \ref{FIG-5} for $^{28}$Si+$^{42}$Nd system (see text for the details). The experimental data is taken from Ref. \cite{Sinha1997}.}	
\end{figure}

\begin{figure}
	\includegraphics[width=8.8cm]{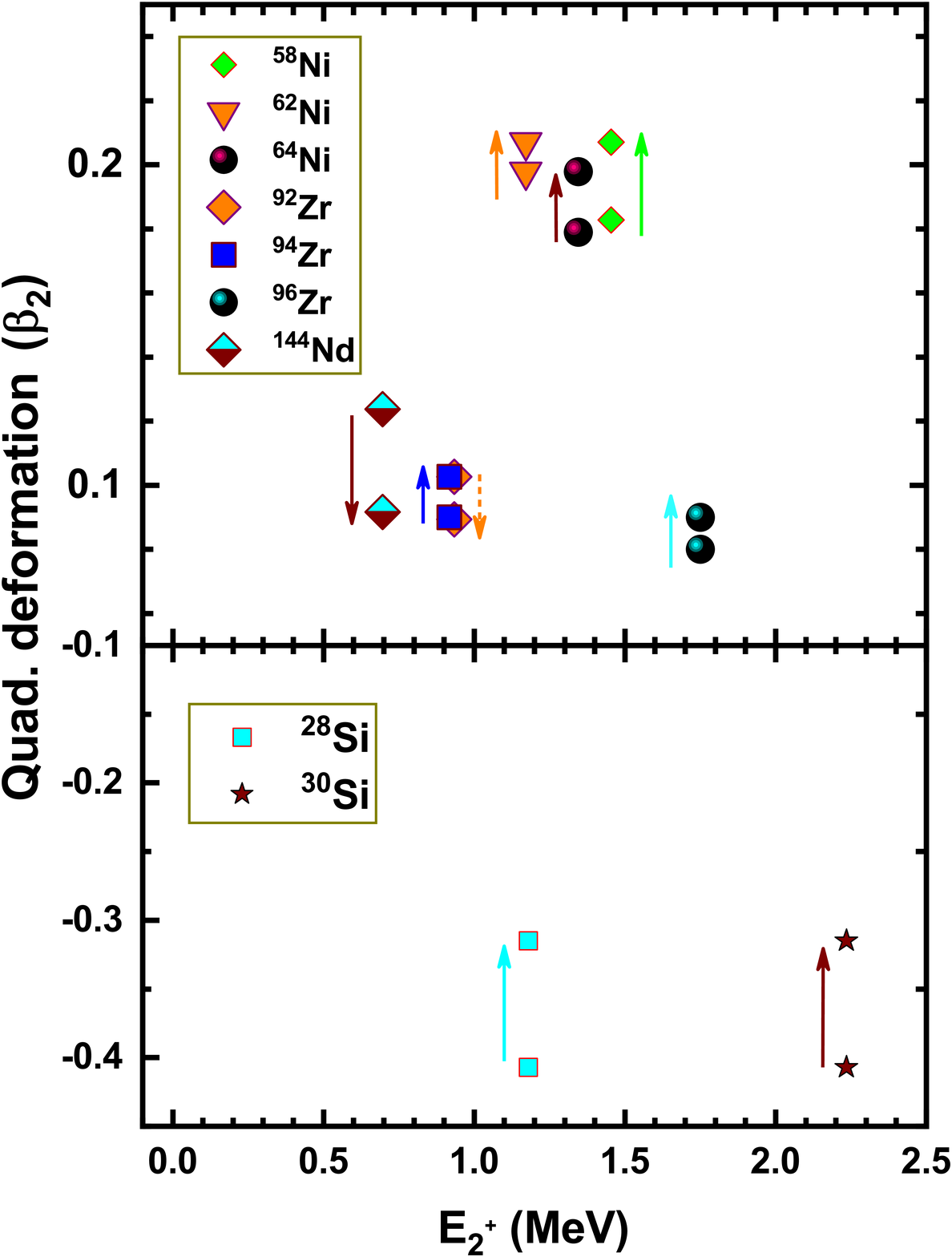}
	\caption{\label{FIG-9} Quadrupol deformation ($\beta_{2}$) as a function of 2$^{+}$ state energy of different nuclei for the systems which have +Q$_{2n}$ channel. The solid lines with the arrow indicate the increment and decrement in deformation after the 2n transfer.  }	
\end{figure}

Furthermore, a comparison is made between the experimental data \cite{Kalkal2010,Khushboo2017} and ECC predictions for $^{28}$Si+$^{94,96}$Zr reactions, having positive Q values up to four and six-neutron pickup transfer channels, respectively (see Table \ref{table2}).
In the ECC model, neutron transfer channels with positive $Q$-values, along with collective excitations in participating nuclei, are taken into account one after the other (see Fig. \ref{FIG-7}(c,d)). It can be observed from the same figures that ECC predictions with up to 2$n$ transfer are reasonally explaining the experimental fusion data throughout the energy range. The aforementioned argument supports the conclusion that the role of up to 2$n$ transfer with Q $>$ 0 on sub-barrier fusion is significant for these reactions, and further inclusion of 3$n$ and 4$n$ transfer channels did not help much, which describes the no role of multi neutron transfer (more than 2$n$) on fusion data in these systems. Therefore, the curves for 1-3n and 1-4n are coinciding with 1-2n transfer channel calculations.
Similar observaions was also reported for $^{28}$Si+$^{158}$Gd \cite{Prajapat2023}, $^{40}$Ca+$^{70}$Zn \cite{Khushboo2019}, and $^{32}$S+$^{90,96}$Zr \cite{Zhang2010} systems recently.
Also, it is good to mention that transfer probability decreases with an increasing number of neutron transfers. Hence, a significant effect is visible only for one or two neutron transfer channels.
One can also see that enhancement in fusion cross-section is significantly large for $^{28}$Si+$^{96}$Zr as compared to $^{28}$Si+$^{90}$Zr, signifying the crucial role of two-neutron pickup transfer with positive Q-value in the first system.

\begin{figure*}
	\begin{subfigure}
		\centering
		\includegraphics[width=16.0cm]{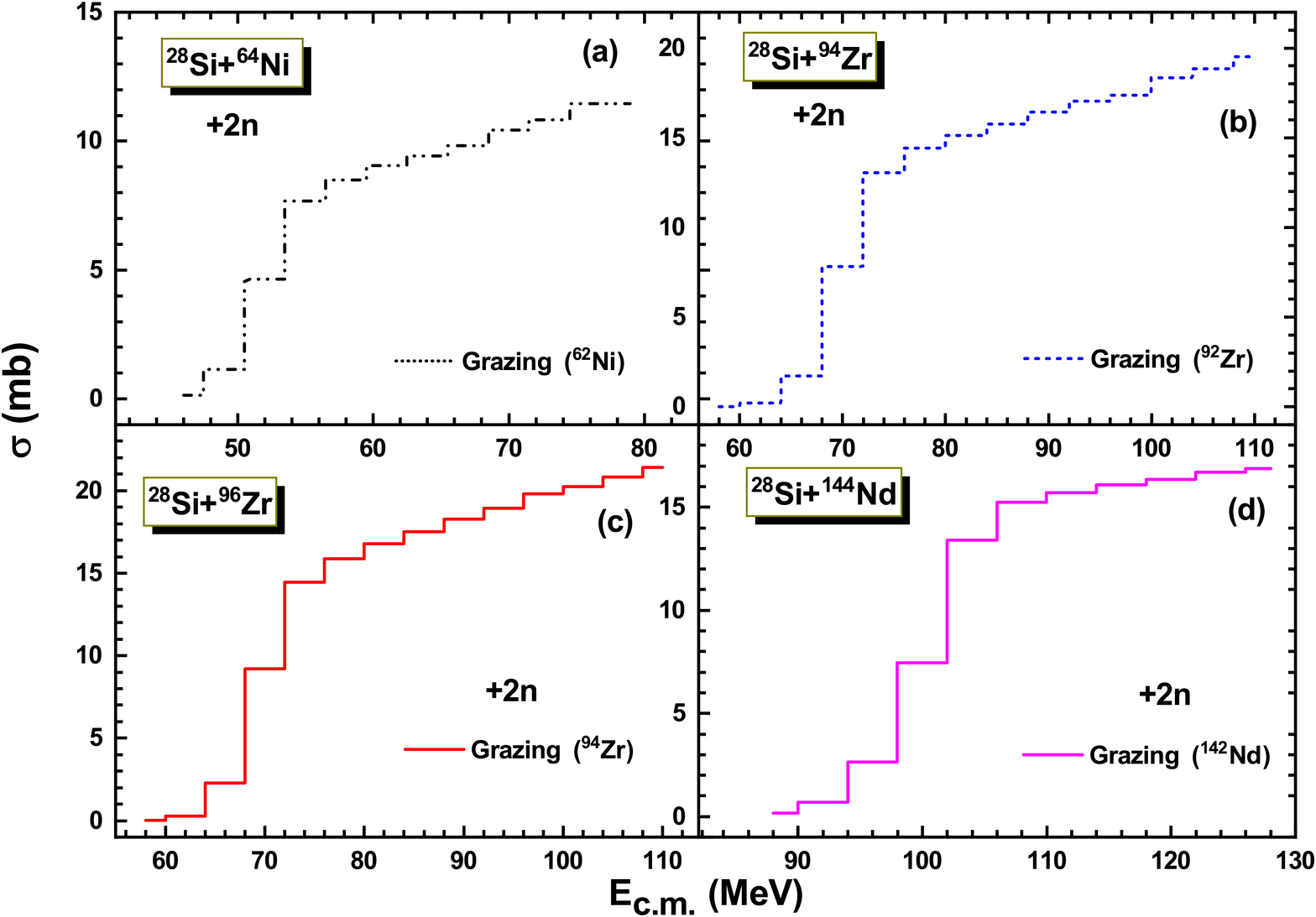}
	\end{subfigure}
	\caption{\label{FIG-10} Grazing predictions for the target-like-ions produced after 2n pickup by projectile from target for (a) $^{62}$Ni from $^{28}$Si+$^{64}$Ni, (b) $^{92}$Zr from $^{28}$Si+$^{94}$Zr, (c) $^{94}$Zr from $^{28}$Si+$^{96}$Zr, and (d) $^{142}$Nd from $^{28}$Si+$^{144}$Nd reaction.}	
\end{figure*}

\subsection{$^{28}$Si+$^{144}$Nd}

To decipher further the impact of multi-neutron transfer on sub-barrier fusion dynamics, a relatively heavy (Z$_{p}$Z$_{t}$ = 840) system with the same projectile $^{28}$Si and a spherical target $^{144}$Nd is chosen with positive Q-values of up to four neutron pickup transfer (see Table \ref{table2}). The experimental data has been taken from Ref. \cite{Sinha1997}. In their studies \cite{Sinha1997}, the coupled channel approach in CCNSC/CCMOD code was broken down when the 2n transfer couplings and higher order excitations in colliding partners were considered, and the experimental data could not be explained.
Thus, the empirical channel coupling calculations are performed by considering no-coupling (uncoupled), without transfer, and with transfer channels, as shown in Fig. \ref{FIG-8}. One may observe that the main effect in fusion enhancement comes from 1n+2n channels on sub-barrier fusion, whereas further inclusion of transfer channels (3n and 4n) does not affect the fusion enhancement probability. This may occur because coupling to the neutron transfer channel with positive Q-values only significantly affects the fusion probability if the process of neutron exchange happens before overcoming the Coulomb barrier.

Thus, it can be concluded that the transfer of only a few neutrons (one or two) influences the transfer probability to a great extent, whereas multi-neutron ($>$2) transfers do not alter it much.

The sub-barrier fusion cross-sections are quite sensitive to the $\beta_{2}$ of participating nuclei. Therefore, we have checked the $\beta_{2}$ before and after the 2n transfer for those systems with the PQNT channel, and the deformation parameters are taken from Ref. \cite{Raman2001}. Moreover, Sargsyan \textit{et al.} \cite{Sargsyan2012} proposed that the reactions having +Q$_{2n}$ should show an enhancement in sub-barrier fusion cross-sections if the $\beta_{2}$ of colliding partners increases and the mass asymmetry decreases after 2n transfer. It also pointed out that neutron transfer weakly influences the sub-barrier fusion cross-section if deformations do not alter or slightly decrease. 
Therefore, we have shown the deformation ($\beta_{2}$) of nuclei before and after the 2n transfer for those reactions in which the +Q$_{2n}$ (see Table \ref{table2}) as a function of first 2$^{+}$ energy of nuclei (before 2n transfer). As it can be noticed from the Fig. \ref{FIG-9},  $^{28}$Si+$^{64}$Ni and $^{28}$Si+$^{94,96}$Zr deformation increases after 2n transfer and enhancement in sub-barrier fusion cross-section also observed, follows the systematics made by Sargsyan \textit{et al.} \cite{Sargsyan2012}. Whereas, slight increment in $\beta_{2}$ observed for $^{62}$Ni and no effect of PQNT observed in $^{28}$Si+$^{62}$Ni reaction. Moreover, $^{30}$Si+$^{58}$Ni and $^{28}$Si+$^{144}$Nd do not follow the Sargsyan \textit{et al.} \cite{Sargsyan2012} systematics, similar behaviour was also reported for $^{30}$Si+$^{142}$Ce \cite{Kaur2024}. It is to be noted that mass-asymmetry ($\eta$) decreases for all the systems after the 2n transfer except for the $^{30}$Si+$^{58}$Ni where +Q$_{2n}$ exist for the stripping channel.

\section{\label{s4}Grazing calculations}
GRAZING code \cite{Grazing_code} is designed to study different observables such as mass and charge, and the energy and angular distributions of MNT products in the grazing regions of heavy-ion collisions. It is based on the semiclassical model of Aage Winther \cite{Winther1994} in which the theoretical description is the approximate solution of CC equations governed by the exchange of particles between the nuclei in a mean-field approximation. Furthermore, the collective excitations and a few nucleon transfers between colliding partners are incorporated using the form factors. 
The nucleon transfer probabilities within this model depend on the single particle-level densities of nucleons, which are defined in terms of free model parameters $\delta^{n}$ (for neutrons) and $\delta^{p}$ (for protons). These free parameters can be tuned to reproduce the measured data within the limits of $\delta^{n} \geq$ 5 and $\delta^{p} \leq$ 10, whereas the default values are $\delta^{p}$ = $\delta^{n}$ = 8. In the present calculations, the default and free parameters of collective excitations are being used.

The GRAZING calculations are done for $^{28}$Si+$^{64}$Ni, $^{28}$Si+$^{94,96}$Zr, and $^{28}$Si+$^{144}$Nd reactions where the evidence of 2n transfer observed. Thus, the cross-section of target-like fragments, e.g., $^{62}$Ni, $^{92,94}$Zr, and $^{142}$Nd, which are produced after 2n pickup by their respective projectiles from targets is calculated and shown in Fig.  \ref{FIG-10}(a-d), respectively.
The predicted cross-sections are grossly close to the difference between those with (2n) and without neutron transfer predictions by the ECC model. This reflects the reliability of these calculations with two different models.
Moreover, the predicted cross-sections of TLFs from the above-mentioned reactions can also be used as an input to measure them experimentally. 
Furthermore, such an advanced understanding of the MNT characteristics of such stable nuclei can be applied to low-intensity radioactive secondary beams \cite{Michimasa2014}.

\begin{table*}
	\caption{\label{table3}Experimentally extracted and theoretically \cite{Kumari2015,Zhang2016,Gharaei2019,Ghodsi2013,Dutt2010,Bass1977} estimated nuclear potential parameters for different selected systems.}
	\begin{ruledtabular}
		\begin{center}
			\renewcommand{\arraystretch}{0.7}
			\begin{tabular}{ccccccccc}
				
				\multirow{1}{2.2cm}{\bfseries Reactions }
				& \multicolumn{1}{c}{ Expt.  }
				& \multicolumn{1}{c}{Prox 77} 
				& \multicolumn{1}{c}{Prox 88} 
				& \multicolumn{1}{c}{Bass 80}
				& \multicolumn{1}{c}{Prox 2000}
				& \multicolumn{1}{c}{Prox 2010}
				& \multicolumn{1}{c}{Zhaung et al.}
				& \multicolumn{1}{c}{Kumari et al.} \\
				
				%	&\multicolumn{6}{c} {  \bfseries Cross section (mb) }\\
				\hline 
				
				& \multicolumn{1}{c}{V$_{b}$, R$_{b}$  } 
				& \multicolumn{1}{c}{V$_{b}$, R$_{b}$  }
				& \multicolumn{1}{c}{V$_{b}$, R$_{b}$  }
				& \multicolumn{1}{c}{V$_{b}$, R$_{b}$ }
				& \multicolumn{1}{c}{V$_{b}$, R$_{b}$ }
				& \multicolumn{1}{c}{V$_{b}$, R$_{b}$ } 
				& \multicolumn{1}{c}{V$_{b}$, R$_{b}$ } 
				& \multicolumn{1}{c}{V$_{b}$, R$_{b}$ } \\
				
				\hline
				$^{28}$Si + $^{58}$Ni &53.1, 8.1   & 55.6, 9.3 &  54.3, 9.6& 53.0, 9.8 & 54.5, 9.6 & 54.1, 9.6 & 51.8, 10.0 & 53.2, 9.7  \\
				
				$^{28}$Si + $^{62}$Ni & 51.6, 8.2   & 55.0, 9.4  & 53.7, 9.7 &52.4, 9.9 & 53.8, 9.7 & 53.5, 9.7 & 51.5, 10.1 &52.5, 9.7 \\
				
				$^{28}$Si + $^{64}$Ni &  51.0, 8.1   & 54.7, 9.5 & 53.4, 9.7 & 52.1, 10.0 & 53.4, 9.8 & 53.2, 9.8 & 51.4, 10.1  & 52.2, 9.8 \\
				
				$^{28}$Si + $^{90}$Zr & 70.9, 8.8 & 74.7, 10.0 & 73.1, 10.2 & 71.3, 10.4 & 73.0, 10.3 & 72.7, 10.3 & 70.2, 10.6 & 70.8, 10.2 \\
				
				$^{28}$Si + $^{92}$Zr & 70.1, 9.5& 74.4, 10.0 & 72.8, 10.3 & 71.1, 10.5 & 72.6, 10.3 & 72.4, 10.3 & 70.1, 10.6 & 70.5, 10.2 \\
				
				$^{28}$Si + $^{94}$Zr & 70.2, 8.8 & 74.1, 10.1 & 72.5, 10.3 & 70.8, 10.5 & 72.3, 10.4 & 72.1, 10.4 & 69.9, 10.6 & 70.1, 10.3\\
				
				$^{28}$Si + $^{96}$Zr & 69.8, 9.8 & 73.8, 10.1 & 72.2, 10.3 & 70.5, 10.6 & 71.9, 10.4 & 71.8, 10.4 & 69.8, 10.6 & 69.8, 10.3\\
				
				$^{28}$Si + $^{144}$Nd & 99.9, 10.2 & 104.1, 10.8 & 102.1, 11.0 & 99.8, 11.2 & 101.6, 11.1 & 101.5, 11.1 & 98.7, 11.3 & 98.0, 10.9 \\
				
				$^{30}$Si + $^{58}$Ni & 52.6, 8.6& 55.1, 9.4 & 53.8, 9.7 & 52.5, 9.9 & 53.9, 9.7 & 53.6, 9.7 & 51.6, 10.1& 52.7, 9.7 \\
				
				$^{30}$Si + $^{62}$Ni & 51.7, 9.5 & 54.5, 9.5 & 53.2, 9.8 & 51.9, 10.0 & 53.1, 9.8 &53.0, 9.8 & 51.4, 10.1 & 52.0, 9.8 \\
				
				$^{30}$Si + $^{64}$Ni & 51.6, 10.1 & 54.2, 9.6 & 53.0, 9.8 & 51.6, 10.1 & 52.7, 9.9 & 52.7, 9.9 & 51.3, 10.1 & 51.7, 9.8 \\				
				
			\end{tabular}
		\end{center}
	\end{ruledtabular}
\end{table*}  
%\end{table}  

\begin{figure*}
	\begin{subfigure}
		\centering
		\includegraphics[width=16.0cm]{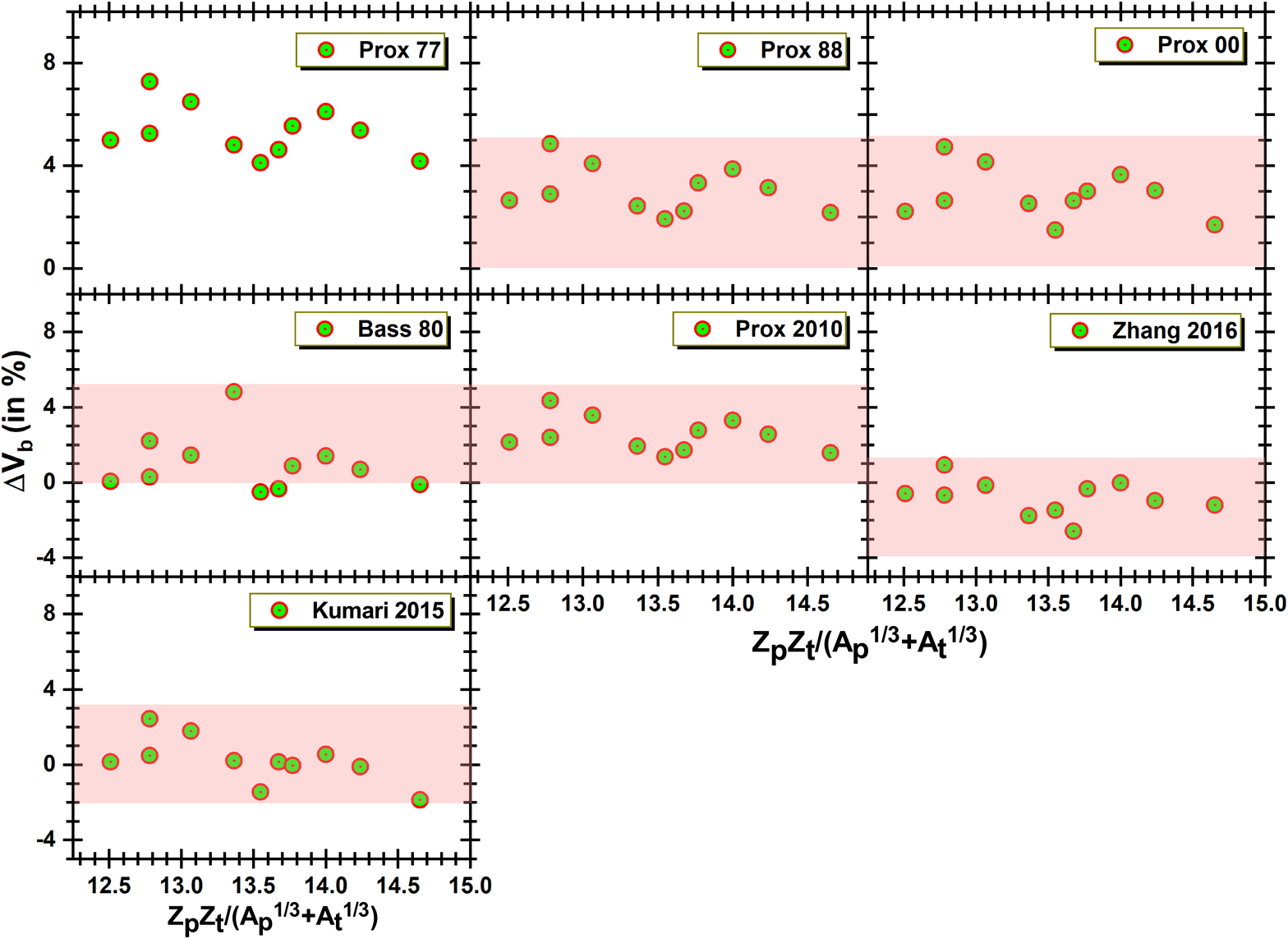}
	\end{subfigure}
	\caption{\label{FIG-11}  The percentage difference $\Delta V_{b}$($\%$) between theoretical and experimental fusion barrier heights as a function of Z$_{p}$Z$_{t}$/(A$_{p}^{1/3}$+A$_{t}^{1/3}$). The shaded area represents those potentials where the deviation is within 5 $\%$.}	
\end{figure*}

\section{\label{s5}Fusion barrier parameters}
The estimation of fusion observables such as barrier height $V_{b}$ and radius $R_{b}$ are the substantial quantities that benchmark the reliability of the theoretical models and assure the quality of measured data. These experimental barrier parameters are derived using experimental fusion data from literature \cite{Stefanini1984,Kalkal2010,Khushboo2017,Sinha1997} at the above barrier energies by comparing the derived slopes and intercepts (from $\sigma_{fus}$ vs. 1/$E_{c.m.}$ relation) with Eq. \ref{eq3}.
These potential barriers are the admixture of the long-ranged repulsive Coulomb term and a short-ranged attractive nuclear term. Thus, on the theoretical front, we calculated the nuclear part of the ion-ion potential using different proximity potentials such as Prox 77, Prox 88, Prox 00, Prox 10, etc., and the addition of Coulomb term gives the total interaction potential $V(r)$ (in MeV). 

%Afterward, the fusion barrier parameters were extracted using the following conditions  [Eq. \ref{eq11}].
%\begin{equation}
%	\label{eq11}
%	\frac{dV(r)}{dr}\Big\vert_{r=R_{b}} = 0, \hspace{0.5cm}\text{and} \hspace{0.5cm} \frac{d^{2}V(r)}{dr^{2}}\Big\vert_{r=R_{b}} \leq 0
%\end{equation}

Afterward, the fusion barrier parameters were extracted using the using the pocket formulas of different versions of proximity potentials; Prox 77, Prox 88, Prox 00, Prox 10 \cite{Gharaei2019,Dutt2010,Ghodsi2013}, Bass potential \cite{Bass1977}, and two parametrized forms;   Kumari \textit{et al.} \cite{Kumari2015}, Zhang \textit{et al.} \cite{Zhang2016}, barrier parameters were calculated and compared with the experimental parameters as listed in Table \ref{table3}.

Furthermore, to access the quality and predictive power of different forms of potentials, we determine the percentage difference of fusion barrier height ($\Delta V_{b}$) as a function of Z$_{p}$Z$_{t}$/(A$_{p}^{1/3}$+A$_{t}^{1/3}$), where Z$_{p}$, Z$_{t}$ and A$_{p}$, A$_{t}$ are the atomic and mass numbers of projectile and target, respectively, defined as following:

\begin{equation}
	\label{eq11}
	\Delta V_{b} (\text{in} \hspace{0.1cm} \%) = \frac{V_{b}^{theor}-V_{b}^{expt}}{V_{b}^{expt}} \times 100
\end{equation}

where V$_{b}^{expt}$ and V$_{b}^{theor}$ are the experimentally derived and theoretically calculated fusion barrier height. One can observe from Table \ref{table3} that the different potentials are in good agreement within 5 $\%$ (see Fig. \ref{FIG-11}) difference with the experimentally derived barrier height for all 11 systems. However, Prox 77 potentially reproduces the experimental barrier height within a 10 $\%$ difference, and we do not find any correlation with the asymmetry parameter of the involved reactions.

Similarly, the $\%$ difference for barrier position was determined, and it was found that theoretical models can reproduce barrier positions within 20 $\%$ deviations. This can be due to the fact that there is enormous variation in extracting barrier positions experimentally. Also, it has been realized that certain factors such as the addition or removal of neutrons \cite{Nicolis2004,Puri1998} and deformation (oblate/prolate) \cite{Denisov2007} in colliding nuclei change the fusion barrier position. This comparison between experimental and theoretical data could be helpful in further refinements in different parameterized forms of $V_{b}$ and radius $R_{b}$. Also, it can be used to predict theoretical fusion cross-sections.

\section{\label{s6} Conclusion}

The influence of a few neutron transfer channels with Q $>$ 0 on sub-barrier fusion cross-sections is studied for several systems within the ECC model framework. A good agreement between model calculations and experimental data is achieved. It is found that experimental data is well reproduced for $^{28}$Si+$^{62}$Ni, $^{30}$Si+$^{58,62,64}$Ni, and $^{28}$Si+$^{90}$Zr systems by considering inelastic excitations such as rotational coupling in $^{28}$Si and vibrational couplings in $^{30}$Si and target nuclei involved in these reactions. However, the fusion data can be explained only by incorporating the neutron transfer channels along with inelastic excitations for $^{28}$Si+$^{64}$Ni, $^{28}$Si+$^{92,94,96}$Zr, and $^{28}$Si+$^{144}$Nd reactions which have the PQNT channels. Therefore, it can be concluded that the role of up to 2n transfer is sufficient to explain the fusion data for such reactions, and neutron transfer ($ >$ 2n) is found insignificant despite having PQNT channels. Furthermore, the GRAZING calculations were performed to estimate the cross-section of target-like fragments after 2n transfer for those reactions in which the role of 2n is found significant. These calculations are also foreseen for MNT and quas-elastic experiments in the near future.

The fusion barrier parameters, such as barrier height and radius, were derived using the experimental data from literature for several reactions and compared with different potential models. It is found that all potential models are in good agreement within 5 $\%$ difference with the experimentally derived barrier height for all 11 systems. However, Prox 77 potentially reproduces the experimental barrier height within a 10 $\%$ difference.
The $\%$ difference for barrier position was 20 $\%$ between model predictions and experiment.

\end{document}